**Laser-induced crystallization of copper oxide thin films: A comparison made between Gaussian and chevron-beam profiles provides a clue for the failure of Gaussian-beam profile**

Willam Bodeau[a], Kaisei Otoge[b], Wenchang Yeh[b*], Nobuhiko P. Kobayashi[a,c,*]

[a]Electrical and Computer Engineering Department, Baskin School of Engineering, University of California Santa Cruz, Santa Cruz, California, United States, 95064

[b]Graduate School of Natural Science and Technology, Shimane University, Matsue, Shimane, Japan, 690-8504

[c]Nanostructured Energy Conversion Technology and Research (NECTAR), University of California Santa Cruz, Santa Cruz, California, United States, 95064

*Corresponding authors

**Abstract.** The use of laser with a Gaussian-beam profile is frequently adopted in attempts of crystallizing non-single-crystal thin films; however, it merely results in the formation of poly-crystal thin films. In this paper, selective area crystallization of non-single-crystal copper(II) oxide (CuO) is described. The crystallization is induced by laser – laser-induced crystallization – with a beam profile in the shape of chevron. The crystallization is verified by the exhibition of a transition from a non-single-crystal phase consisting of small (~100 nm x 100 nm) grains of CuO to a single-crystal phase of copper(I) oxide ($Cu_2O$). The transition is identified by electron back scattering diffraction and Raman spectroscopy, which clearly suggests that a single-crystal domain of $Cu_2O$ with size as large as 5 μm x 1 mm develops. Provided these experimental findings, a theoretical assessment based on a cellular automaton model, with the behaviors of localized recrystallization and stochastic nucleation, is developed. The theoretical assessment can qualitatively describe the laser beam geometry-dependence of vital observable features (e.g., size and gross geometry of grains) in the laser-induced crystallization. The theoretical assessment predicts that differences in resulting crystallinity – either single-crystal or poly-crystal – primarily depend on a geometrical profile with which melting of non-single-crystal regions takes place along the laser scan direction; concave-trailing profiles yield larger grains which lead to single-crystal while convex-trailing profiles results in smaller grains which lead to poly-crystal, casting light on the fundamental question *Why does a chevron-beam profile succeed in producing single-crystal while a Gaussian-beam profile fails?*

**1. Introduction**

Laser-induced crystallization (LIC) frequently adopted in attempts of crystallizing non-single-crystal (NSC) thin films offers attractive features advantageous for functional devices that need to be built on NSC substrates such as glasses for which epitaxial growth – a conventional technique to obtain single-crystal (SC) semiconductor thin films – does not serve. LIC has a long history, dating back to the late 1970s[1,2,3,4,5] with a significant emphasis on elementary semiconductor thin films containing a single chemical element such as Si to develop thin film transistors. LIC was also exploited, to a smaller extent, for semiconductor thin films containing multiple chemical elements (e.g., group IV compound semiconductors, group III-V compound semiconductors, metal oxide semiconductors[6,7,8,9,10,11,12]). However, in all these cases of both elementary and alloy semiconductor thin films, femtosecond laser or excimer laser with a Gaussian-beam profile was used dominantly, merely producing thin films made of crystalline domains with nominal lateral size on the order of 1 µm; in another way of saying, it failed to produce SC domains that provide usable areas large enough for practical devices. In this paper, selective area crystallization of NSC copper(II) oxide (CuO) is described. The crystallization is achieved by LIC with a beam profile in the shape of chevron – a marked contrast to a Gaussian-beam profile. The crystallization is verified by observing a transition from a NSC phase consisting of small (~100 nm x 100 nm) grains of CuO to a SC phase of copper(I) oxide ($Cu_2O$). The transition is identified by electron backscattering diffraction (EBSD) and Raman spectroscopy, clearly suggesting that a single-crystal domain of $Cu_2O$ with size as large as 5 µm x 1 mm develops. Provided this experimental demonstration, a theoretical assessment based on a cellular automaton model, with the behaviors of localized recrystallization and stochastic nucleation, is developed. The theoretical assessment qualitatively predicts the dependence of vital observable features (e.g., size and gross geometry of domains) obtained in the experiment. The theoretical assessment further predicts that differences in resulting crystallinity – either SC or poly-crystal – primarily depend on the geometric details with which non-single-crystal regions exposed to laser melt in relation to the scan direction of the laser. Concave-trailing profiles yield larger grains which lead to SC while convex-trailing profiles result in smaller grains which lead to poly-crystal, casting light on the fundamental question *Why does a chevron-beam profile succeed in producing SC while a Gaussian-beam profile fails?*

## 2. Experiment

The concept of LIC with a chevron-beam profile – chevron-beam LIC – is depicted in Fig. 1. Fig. 1(a) illustrates the initial structure that consists of a NSC layer deposited on a NSC substrate. The NSC layer is covered with a cap layer. In our implementation, the cap layer, the NSC layer, and the NSC substrate were a 200 nm thick silicon dioxide ($SiO_2$) layer, a 130 nm thick CuO layer, and a fused-silica substrate, respectively. The CuO layer and the $SiO_2$ cap layer were deposited sequentially in a single vacuum chamber without breaking the vacuum by DC magnetron sputtering and pulsed DC magnetron sputtering, respectively, at room temperature. A specific thickness of 130 nm was chosen for the CuO layer to ensure sufficient absorption of the laser with the wavelength of 405 nm. The presence of the $SiO_2$ cap layer was found to be critical to reducing incongruent evaporation from the underlying CuO layer during chevron-beam LIC. As schematically depicted in Fig. 1(b), a laser with a chevron-beam profile travels through the $SiO_2$ cap layer and locally melts the CuO layer, subsequently the melted region becomes SC upon solidification as illustrated by the cross-sectional region in red marked "SC". Fig. 1(c) displays a top view of the SC strip in red (The cap layer on the SC strip is not shown to reveal the SC strip for a display purpose). The laser with a chevron-beam-profile is depicted by two green lines joined at their ends. The black rightwards arrow represents the direction along which the laser is scanned. The cross-sectional region of the SC strip in Fig. 1(b) is drawn along the black dotted line in Fig. 1(c). $W$ ~10 μm in Fig. 1(b) and $L$ in Fig. 1(c) are the width and length of the SC strip. $W$ is comparable to the opening of the chevron-beam profile in green in Fig. 1(c) while $L$ is only limited by the linear translational motion of the moving stage (i.e., the distance over which the laser is scanned) and can be extended as needed. The chevron-beam profile was established by having the output of a 405 nm wavelength multimode CW laser-diode (LD) pass through a one-sided dove prism that converted the line-beam profile into a chevron-beam profile[13] focused on the CuO layer. The initial structure in Fig. 1(a) was mounted on a linearly moving stage that advanced at a speed (i.e., scan rate $R_{scan}$) in the range of 0.4 - 5 mm/s with respect to the fixed position of the laser with the laser power output $P_L$ varied in the range of 50 - 140 mW; thus, provided the geometrical dimension of the chevron-beam profile, areal laser power density was varied in the range of 1 - 1.5 x $10^6$ W/cm$^2$. EBSD was performed after removing the SiO$_2$ cap layer to identify the phase transition from NSC CuO to SC Cu$_2$O and Raman spectroscopy was carried out with an excitation wavelength of 514.5 nm at room temperature to evaluate crystallinity of the resulting SC strips.

Fig. 2(a), (c), and (f) show images of three strips collected by scanning electron microscopy (SEM) by titling the strips by 70 degrees. The three strips were prepared by the chevron-beam LIC with three different $P_L$ in panels (a) 58.1 mW, (c) 61.2 mW, and (f) 64.3 mW ($R_{scan}$ was fixed at 0.4 mm/s). The black arrow in Fig. 2(a) represents the scanning direction of the laser for all the three strips (i.e., for all the panels in Fig. 2). Correspondingly, crystal orientation maps, obtained by EBSD to reveal a specific crystal orientation in parallel to the surface normal of the three strips, are displayed in Fig. 2(b), (d), and (g) for the strips prepared at $P_L$ = 58.1 mW, 61.2 mW, and 64.3 mW, respectively. The triangular color map is in reference to the major crystal orientations. For the strips prepared at $P_L$ = 61.2 mW and 64.3 mW shown in Fig. 2(d) and (g), respectively, correlative grain boundary maps, also collected by EBSD, are presented in Fig. 2(e) and (h). A grain boundary map is not provided for the strip prepared at $P_L$ = 58.1 mW in Fig. 2(b) because this strip is found to be poly-crystal for which the presence of grain boundaries is apparent even in its crystal orientation map in Fig. 2(b). A series of the SEM images in Fig. 2(a), (c), and (f) indicates that distinctive surface features associated with the formation of strip by the chevron-beam LIC amplifies in the direction parallel to the scan direction of the laser as $P_L$ increases. Moreover, the surface features appear more pronounced as $P_L$ is raised, suggesting that the volume within which laser and the CuO layer interact, both optically and thermally, increases as $P_L$ is raised. The crystal orientation map in Fig. 2(b) indicates the strip prepared at $P_L$ = 58.1 mW contains numerous grain boundaries running across the strip, resulting in a poly-crystal as a whole with the presence of domains with color that changes discontinuously along the length of the strip. While the strip prepared at $P_L$ = 58.1 mW in Fig. 2(a) contains randomly oriented grains, the crystal orientation maps of the strips prepared at $P_L$ = 61.2 mW and 64.3 mW in Fig. 2(d) and (g), respectively, consist of a single domain (i.e., these domains are SC) although the domains seem to rotate along the axes parallel to the length of the strips, which is seen as a continuous change in color along the length. The grain boundary maps in Fig. 2(e) and (h) further confirm that these trips are SC as no continuous random angle grain boundaries (RGB) nor coincident site lattice boundaries (CSLB) are found. It is worth mentioning that when $P_L$ was raised to 67.4 mW, voids appeared within a strip although the strip remained SC.

Fig. 3 shows a series of Raman spectra collected from the strips prepared using various $P_L$ in the range of 38 – 138 mW ($R_{scan}$ was set to 1 mm/s for all these strips). The spectra of strips prepared at $P_L$ in the range of 53 – 138 mW all show the dominant phonon mode at 218 cm$^{-1}$ that represents the second overtone of the phonon mode at 109 cm$^{-1}$ – an inactive Raman mode that is only infrared-allowed in superb $Cu_2O$ crystals[14] – indicating that the strips prepared at $P_L$ in the range of 53 – 138 mW bear structural integrity comparable to $Cu_2O$ formed under conditions near thermal equilibrium[15]. Characteristic phonon modes associated with CuO (e.g., ~300 cm$^{-1}$) are not seen in these spectra, confirming that the strips prepared at $P_L$ in the range of 53 – 138 mW are predominantly made of crystalline $Cu_2O$[16,17]. The presence of the well-defined second-order overtone at 218 cm$^{-1}$ further indicates that these strips have high crystallographic integrity[21]. In contrast, the spectra of the strip prepared at $P_L$ at 38 and 45 mW evidently lack the $Cu_2O$ phonon mode at 218 cm$^{-1}$. More specifically, the spectrum of $P_L$ = 38 mW shows no distinguishable peaks while the spectrum of $P_L$ = 45 mW shows two peaks at 296.5 cm$^{-1}$ and 341.3 cm$^{-1}$ characteristics to crystalline CuO[18]. For these two peaks, a noticeable tail exists on the side of lower wavenumbers, which indicates that CuO exists in the form of poly-crystal consisting of crystalline grains with a rather wide range of size[19]; smaller grain sizes result in broader peaks and redshift[20,21,22]. Evidently, two phase transitions underwent as $P_L$ was raised from 38 mW to 138 mW: the first transition occurring at 45 mW is associated with a transformation of non-crystalline CuO into crystalline CuO and the second transition taking place at 53 mW is associated with a transition from crystalline CuO to crystalline $Cu_2O$. Although complex oxidation kinetics of copper at elevated temperatures resulting in the interplay between the two phases, CuO and $Cu_2O$ would contribute to how Raman spectra appear[23,24], the observed modes may largely be attributed to Raman selection rules lifted due to point defects such as Cu vacancies commonly present in p-type $Cu_2O$[25]. Nevertheless, the use of higher $P_L$ promotes the tendency of converting non-crystalline CuO into crystalline $Cu_2O$ via an intermediate phase of crystalline CuO when $R_{scan}$ is appropriately set.

Shown in Fig. 4 are crystal orientation maps of three strips prepared by the chevron-beam LIC with three different $R_{scan}$: panel (a) 10 mm/s, panel (b) 5 mm/s, and panel (c) 1 mm/s ($P_L$ was fixed at 87 mW for all these strips). Fig. 4(a) is filled with pixels of random colors, indicating that the strip prepared with $R_{scan}$ = 10 mm/s exhibits no preferential crystal planes and is deemed non-crystalline. In contrast, the strip in Fig. 4(b) prepared with $R_{scan}$ = 5 mm/s consists of domains of linear size in the range of 1 – 2.5 µm, indicating individual domains grew laterally although these domains are spatially separated by boundaries that appear to be filled with pixels of random colors (i.e., the presence of grain boundaries); thus, the strip in Fig. 4(b) is regarded as poly-crystal. When $R_{scan}$ was decreased to 1 mm/s as in Fig. 4(c), the strip grew into a single domain (i.e., SC) filled primarily with yellowish colors – the middle of [001] and [101] on the reference color map provided for Fig. 2. Fig. 4 affirms that, for a given $P_L$, the use of smaller $R_{scan}$ promotes the growth of a single domain, increasing the chance of developing SC.

In Fig. 5, full-wide-at-half-maximum (FWHM) of the $Cu_2O$ peak that appears in each of the Raman spectra shown in Fig. 3 is plotted as a function of $P_L$ for two different $R_{scan}$: 5 mm/s and 1 mm/s. A few features in common for the two $R_{scan}$ reveal that the FWHM decreases – presumably improvements in crystallinity – almost linearly as $P_L$ is raised, and the FWHM appears to converge at ~10.5 cm$^{-1}$ as $P_L$ exceeds 140 mW regardless of $R_{scan}$. Fig. 5 clearly suggests that, for a given $R_{scan}$, there exits a threshold $P_L$ above which the formation of SC $Cu_2O$ from non-crystalline CuO is energetically preferred. Once the threshold $P_L$ is reached (e.g., ~53 mW for $R_{scan}$ = 1 mm/s as seen in Fig. 3), crystallinity of $Cu_2O$ improves (i.e., FWHM decreases), for a given $R_{scan}$, as $P_L$ increases. The interplay between $R_{scan}$ and $P_L$ may be elucidated by considering the ratio $R$ of $P_L$ to $R_{scan}$, which has a dimension of J/m (i.e., energy per length). For instance, using $R_{scan}$ of 5 mm/s at 76 mW (i.e., $R$ = 15 J/m) results in FWHM of 13.2 cm$^{-1}$ while using $R_{sacn}$ of 5 mm/s at 138 mW (i.e., $R$ = 27 J/m) results in FWHM of 11.7 cm$^{-1}$ which is comparable to that obtained by using $R_{scan}$ of 1 mm/s at 92 mW (i.e., $R$ = 92 J/m). Evidently, a choice of $R_{scan}$ for a given $P_L$ is critical in improving crystallinity, maximizing energy efficiency, and increases the throughput of the chevron-beam LIC; thus, a theoretical assessment was carried out to illustrate the dependence of resulting crystallinity on $R_{scan}$, and more importantly, to address the fundamental question *Why does a chevron-beam profile succeed in producing SC while a Gaussian-beam profile fails?*

### 3. Theoretical assessment

In our efforts of visualizing how the use of a chevron-beam profile and $R_{scan}$ contribute to the formation of SC, the crystallization was modeled with a non-differential cellular automaton (NDCA) evolved on a two-dimensional square grid of cells. Each cell can be configured either in a so-called liquid state (visualized as white) or in one of various solid states, representing variations in crystallographic orientation (visualized as different colors). While a solid cell, once formed, remains solid with a distinctive color (unless re-melted by laser), all liquid cells have a chance to solidify at every time step. A laser beam, with a specific beam profile, scanned over a sample, as illustrated in Fig. 1(b) and (c), is modeled by setting cells located within a high-temperature region defined by the laser to the liquid state, and then the laser is moved at the rate $R_{scan}$. In our two-dimensional model, an initial, NSC thin film is defined as a rectangle region – the top view of a NSC thin film – that consists of many square cells as illustrated in Fig. 6(a) (Only 9 cells are shown in panel (a) for the display purpose.). Initially, a random solid state is assigned to each of the cells in the region, producing an effectively NSC initial condition with no distinguishable grains. Within the region, a set of cells illuminated by a laser (i.e., cells that are liquid) are shown in white, displaying geometrical details of a laser beam profile. As shown in panel (b), during a time step, a liquid cell (e.g., the center cell in panel (b)) randomly chooses one of 8 surrounding cells and inspects the state of the chosen cell. If the chosen cell is solid, the updating cell changes its state to match that of the chosen cell as shown in panels (c) and (d); this is representative of seeded grain expansion. In contrast, if the chosen cell is liquid as in panel (e), the updating cell either remains liquid as in panel (f) or changes its state to a random solid state as in panel (g), which is governed by the parameter $p_n$ – the probability that local solidification (i.e., nucleation) occurs in a given cell within one time-step. High $p_n$ establishes the condition under which nucleation is preferred while low $p_n$ signifies nucleation is not preferred; thus, a $p_n$ of 0 has solidification only occurring via interface expansion, and a $p_n$ of 1 has all liquid immediately solidifying, similar to the amorphous starting conditions. The trajectory of laser scan used in the modeling is linear, which is characterized by laser scan speed $v_{scan}$ that has a unit of cell numbers per timestep. Because growth is limited to one lattice point per timestep, a seeded growth velocity is implicit in the model, on the order of 0.3 cells per time step.

Figs. 7(a)-(d) show the dependence of the formation of grains on relative $v_{scan}$. Using a higher $v_{scan}$ as in Fig. 7(a) leads to an apparent decrease in grain size, causing an increasingly thick outer portion made

of smaller grains randomly oriented one another. In contrast, as seen in Figs. 7(b) and (c), a reduction in $v_{scan}$ promotes the growth of domains much larger than grains seen in Fig. 7(a), and, eventually, a single domain (i.e., SC) forms as $v_{scan}$ is further reduced as shown in Fig. 7(d), which qualitatively suggests that the solid front fails to keep up with the laser being continuously scanned, and a liquid wake begins to form when $v_{scan}$ increases, promoting nucleation and decreasing grain size.

Two cases evaluated for three types of beam profiles – chevron used in the present study and Gaussian in conventional LIC – are compared in Figs. 8(a) and (b). As described above, the areas in white illustrate the two types of beam profiles. The results qualitatively suggest that the Gaussian-beam profile is likely to fail in producing large domains, while the chevron-beam profile offers a better chance of forming SC, which is consistent with our experiment. Detailed analysis of the trailing edge of the Gaussian-beam profile reveals that, as liquid solidifies, grain boundaries progress vertically with respect to the liquid wake, curving forward and in towards the center. In addition, fresh liquid cells were found to be immediately exposed to the un-melted non-crystalline region, causing growth to start as many small randomly oriented grains growing into the channel from the sides, which clearly suggests that any beam profiles presenting a convex tailing liquid wake will result in numerous randomly oriented grains. This was found to be valid to beam profiles generally characterized by convex curvature as exemplified in Fig. 8(c) that displays an ellipse-beam profile. As far as we are aware, these results explicitly indicate, for the first time, that LIC with a Gaussian-beam profile fails to produce SC, which is observed in many experimental results of conventional LIC. In contrast, the use of a chevron-beam profile dramatically increases nominal grain size. Detailed analysis reveals that the concave shape of the liquid-solid interface shields the newly created liquid, exposing it only to the recently crystallized region immediately behind it.

The dependence of solidification on laser scan direction was also quantitatively examined for various beam profiles. Three types of beam profiles, chevron in Fig. 9(a) and (b), cross in Fig. 9(c) and (d), and ellipse in Fig. 9(e) and (f), were examined. In these panels, white arrows indicate the direction along which the laser beam was scanned; for instance, in Fig. 9(a), a chevron-beam profile was scanned from the bottom to the top (i.e., scan angle $\theta_{scan}$ = 90 degrees) while in Fig. 9(b) a chevron-beam profile was scanted from the left to the right (i.e., $\theta_{scan}$ = 0 degree). Fig. 9(a) and (b) show results of using the chevron-beam profile. The two line segments define a chevron-beam profile with a vertex, in other words, two line segments outline a triangular section with an apex. A poly-crystal results both in the triangular section marked with α and in the area denoted with β when the base of the triangular section is parallel to the scan direction as in Fig. 9(a). Furthermore, the poly-crystal sections are made of swaths with various colors (i.e., various crystal orientations) that appear to have developed in the direction perpendicular to the line segments. In contrast, a SC forms when the base of the triangular section γ is perpendicular to the scan direction as seen in Fig. 9(b), indicating the presence of significant anisotropy in terms of resulting crystallinity that substantially depends on the scan direction with respect to the geometry of the beam profile. Fig. 9(c) and (d) show results of using the cross-beam profile. Two crossing line segments define four triangular sections with a common vertex located at the cross point of the cross-beam profile. In both Fig. 9(c) and (d), a SC forms in the triangular section δ and ε with the base perpendicular to the scan direction, which is consistent with Fig. 9(b). The two triangular sections ζ and η in Fig. 9(c) and θ and ι in Fig. 9(d) with their bases parallel to the scan direction yield poly-crystal made of swaths with various colors, which is consistent with α region in Fig. 9(a). However, these swaths are eventually converted into SC as the trailing triangular sections δ and ε run over them. Unlike the chevron and the cross-beam profiles, the ellipse-beam profiles in Fig. 9(e) and (f) fail to produce a single crystal with $\theta_{scan}$ = 0 and 90 degrees, which is consistent with Fig. 8(c). Nevertheless, poly-crystal domains that appear in Fig. 9(e) seem to be much smaller than those seen in Fig. 9(f), clearly illustrating that the size of poly-crystal domains strongly depends on the curvature of the ellipse – tighter curvatures result in smaller poly-crystal domains.

In Fig. 9(g), performance defined by the size of poly-crystal domains that were averaged over the entire region scanned by laser and were weighted by the total area being scanned by laser is plotted as a function of $\theta_{scan}$ for the three beam profiles: chevron, cross, and ellipse. The three plots colored magenta (chevron), green (cross), and dark blue (ellipse) exhibit unique anisotropy resulting from their specific geometrical relationships between a specific beam profile and a scan direction. As seen in Fig. 9(a) and (b), the chevron-beam profile in magenta exhibits a substantial directional characteristic when the base of the triangular section γ in Fig. 9(b) is perpendicular to the laser scan direction (i.e., $\theta_{scan}$ = 0 degree) while the cross-beam profile in green shows a four-fold rotational symmetry which arises from the fact that the cross-beam profile is merely a shape made by coupling four chevron-beam profiles with a cross point where four apexes of the four triangles meet. In contrast, the ellipse beam profile in dark blue, with a two-fold rotational symmetry associated with the long and the short axes of the ellipse, seems to be substantially inferior to the chevron-beam profile along all the directions, which is consistent with Fig. 9(e) and (f). The collective results shown in Fig. 9 suggest that a critical feature required for the formation of SC is a concave beam shape featuring protective side curves trailing out behind the leading central apex.

Provided advantageous features of the chevron-beam profile, the dependence of the performance on variations in the geometrical factors – length of line segment $L$ expressed as the number of cells and the angle between the two-line segments $\theta$ as illustrated in the inset of Fig. 10 – is examined. In Fig. 10, chevron-beam profiles with $\theta$ smaller than 180° represent a concave-trailing chevron, while $\theta$ larger than 180° represent a convex-trailing chevron, evidently suggesting the importance of the use of a beam profile characterized as concave-trailing to yield SC.

## 4. Summary

The use of laser with a Gaussian-beam profile is frequently adopted in attempts of crystallizing NSC thin films; however, it merely results in the formation of poly-crystal thin films. In this study, selective area crystallization of NSC CuO achieved by LIC with a beam profile in the shape of chevron – chevron-beam LIC with a marked contrast to a Gaussian-beam profile – was demonstrated. The crystallization was verified by observing a transition from a NSC phase of CuO to a SC phase of $Cu_2O$ with size as large as 5 μm x 1 mm. The use of higher $P_L$ raises the tendency of converting non-crystalline CuO into SC $Cu_2O$ via an intermediate phase of crystalline CuO when $R_{scan}$ is appropriately set. Nevertheless, a choice of $R_{scan}$ for a given $P_L$ is critical in improving crystallinity, maximizing energy efficiency, and increasing the throughput of the chevron-beam LIC. Provided the experimental demonstration, a theoretical assessment based on a cellular automaton model was developed, which qualitatively predicts the dependence of vital observable features obtained in the experiment. The theoretical assessment further predicts that differences in resulting crystallinity – either SC or poly-crystal – primarily depend on the geometric details with which NSC regions are exposed to laser melt in relation to the scan direction of the laser. Concave-trailing profiles yield larger grains which lead to SC while convex-trailing profiles result in smaller grains which lead to poly-crystal, casting light on the fundamental question *Why does a chevron-beam profile succeed in producing single-crystal while a Gaussian-beam profile fails?* As far as we are aware, these results explicitly indicate, for the first time, that LIC with a Gaussian-beam profile fails to produce a single crystal, which is observed in many experimental results of conventional LIC. Given advantageous features of the chevron-beam profile, the dependence of the performance on variations in the geometrical factors was examined, evidently suggesting the importance of the use of a beam profile characterized as concave-trailing to yield SC.

**Acknowledgement**

The experimental part of this study, which was done by Shimane University was partially supported by the JST-ASTEP project. The modeling part of this study, which was done by the University of California Santa Cruz was partially supported by the National Science Foundation (U.S.A.) under the award #1562634 (Program manager: Dr. Tom Kuech).

**Figure captions**

**Fig. 1:** The concept of chevron-beam LIC. (a) the initial structure. (b) laser with a chevron-beam profile locally melts the NSC film. Upon cooling, the melted region becomes SC shown as a cross-sectional region in red marked "SC." (c) a top-view of the SC strip in red in (b) (The cap layer on the strip is partially removed to reveal the strip). The laser with a chevron-beam profile is depicted by two green lines joined at their ends. The downward black arrow represents the direction along which the laser is scanned. The dotted line shows a section along which the cross-sectional region in (b) is drawn. $W$ and $L$ are the width and length of the SC strip.

**Fig. 2:** Panels (a), (c), and (f) show images of three strips collected by SEM. The three strips were prepared by the chevron-beam LIC with three different $P_L$: (a) 58.1 mW, (c) 61.2 mW, and (f) 64.3 mW ($R_{scan}$ was fixed at 0.4 mm/s). The black arrow in (a) represents the scanning direction of the laser for all the three strips (i.e., for all the panels in Fig. 2. Panels (b), (d), and (g) show crystal orientation maps the strips prepared at $P_L$ = 58.1 mW, 61.2 mW, and 64.3 mW, respectively, corresponding to panels (a), (c), and (f).

The triangular color map is in reference to the major crystal orientations. For panels (d) and (g), correlative grain boundary maps are presented in panels (e) and (h). Discontinuous random angle grain boundaries (RGB) and coincident site lattice boundaries (CSLB) are show in black and red. The scale bars represent 5 μm.

**Fig. 3:** A series of Raman spectra collected from the strips prepared using various $P_L$ in the range of 38 – 138 mW ($R_{scan}$ was set to 1 mm/s for all these strips). The spectra of strips prepared at $P_L$ in the range of 53 – 138 mW all show the dominant phonon mode at 218 cm$^{-1}$ that represents the second overtone of the phonon mode at 109 cm$^{-1}$ of crystalline $Cu_2O$. In contrast, the spectra of the strip prepared at $P_L$ at 38 and 45 mW evidently lack the $Cu_2O$ phonon mode at 218 cm$^{-1}$. More specifically, the spectrum of $P_L$ = 38 mW shows no distinguishable peaks while the spectrum of $P_L$ = 45 mW shows two peaks at 296.5 cm$^{-1}$ and 341.3 cm$^{-1}$ characteristics to crystalline CuO.

**Fig. 4:** Crystal orientation maps of three strips prepared by the chevron-beam LIC with three different $R_{scan}$: (a) 10 mm/s, (b) 5 mm/s, and (c) 1 mm/s ($P_L$ was fixed at 87 mW for all these strips). (a) the strip filled with pixels of random colors is deemed non-crystalline. (b) the strip that consists of domains of linear size in the range of 1 – 2.5 μm is considered to be poly-crystal. (c) the strip grew into a single domain filled primarily with yellowish colors – the middle of [001] and [101] on the reference color map provided for Fig. 2. The scale bars represent 5 μm.

**Fig. 5**: FWHM of the $Cu_2O$ peak of each spectrum in Fig. 3 is plotted as a function of $P_L$ for two different $R_{scan}$: 5 mm/s and 1 mm/s. For both $R_{scan}$, the FWHM decreases as $P_L$ is raised, and the FWHM appears to converge at ~10.5 cm$^{-1}$ as $P_L$ exceeds 140 mW regardless of $R_{scan}$. Fig. 5 clearly suggests that, for a given $R_{scan}$, there exists a threshold $P_L$ above which the formation of SC $Cu_2O$ from non-crystalline CuO is energetically preferred.

**Fig. 6:** A top view of a NSC thin film that consists of many square cells as illustrated in Fig. 6(a). Initially, a random solid state is assigned to each of the cells in the region, producing an effectively NSC initial

condition with no distinguishable grains. Within the region, a set of cells illuminated by laser (i.e., cells that are liquid) are shown in white, displaying geometrical details of a laser beam profile. (b), during a time step, a liquid cell (e.g., the center cell in panel (b)) randomly chooses one of 8 surrounding cells and inspects the state of the chosen cell. If the chosen cell is solid, the updating cell changes its state to match that of the chosen cell as shown in panels (c) and (d). If the chosen cell is liquid as in (e), the updating cell either remains liquid as in panel (f) or changes its state to a random solid state as in (g).

**Fig. 7:** The dependence of the formation of grains on relative $v_{scan}$. (a) using $v_{scan}$ = 10.0 leads to an apparent decrease in grain size, causing an increasingly thick outer portion made of smaller grains randomly oriented one another. (b) $v_{scan}$ = 1.0 and (c) $v_{scan}$ = 0.7, using a reduced $v_{scan}$ promotes the growth of domains much larger than grains seen (a). (d) using $v_{scan}$ = 0.3 (d) results in a single domain (i.e., SC). For all cases, $p_n$ = 5x10$^{-6}$.

**Fig. 8:** (a) and (b) two cases are evaluated for three types of beam profiles – chevron used in the present study and Gaussian in conventional LIC. The results qualitatively suggest that the Gaussian-beam profile is likely to fail in producing large domains, while the chevron-beam profile offers a better chance of forming SC, which is consistent with our experiment. (c) an ellipse-beam profile also fails to produce a single domain.

**Fig. 9:** The dependence of solidification on laser scan direction was examined for various beam profiles. Three types of beam profiles, chevron in (a) and (b), cross in (c) and (d), and ellipse in (e) and (f) were examined. White arrows indicate the direction along which the laser beam was scanned; for instance, in (a), a chevron-beam profile was scanned from the bottom to the top (i.e., scan angle $\theta_{scan}$ = 90 degrees) while in (b) a chevron-beam profile was scanted from the left to the right (i.e., $\theta_{scan}$ = 0 degree). (g) performance defined by the size of poly-crystal domains that were averaged over the entire region scanned by laser and weighted by the total area being scanned by laser is plotted as a function of $\theta_{scan}$ for the three beam profiles. The three plots colored magenta (chevron), green (cross), and dark blue (ellipse) exhibit

unique anisotropy resulting from their specific geometrical relationships between a specific beam profile and a scan direction.

**Fig. 10:** The dependence of the performance on variations in the geometrical factors – length of line segment *L* expressed as the number of cells and the angle between the two-line segments $\vartheta$ as illustrated in the inset is examined. Chevron-beam profiles with $\theta$ smaller than 180° represent a concave-trailing chevron, while $\theta$ larger than 180° represent a convex-trailing chevron.

FIG1

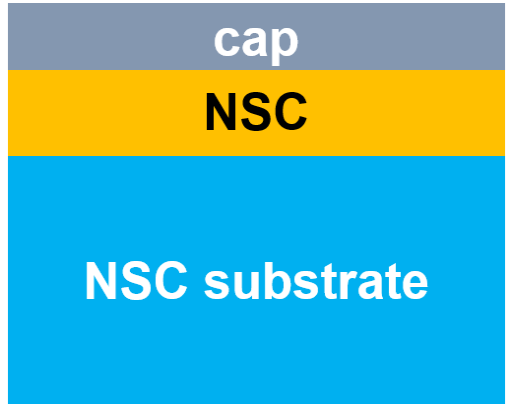

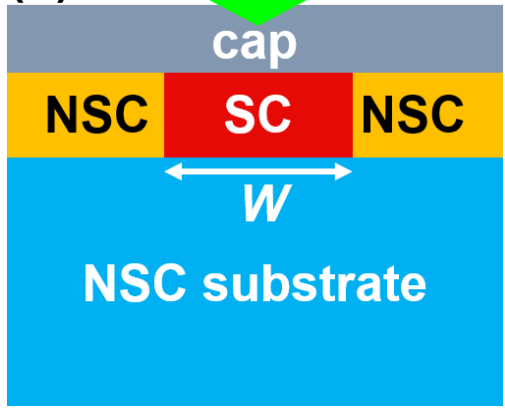

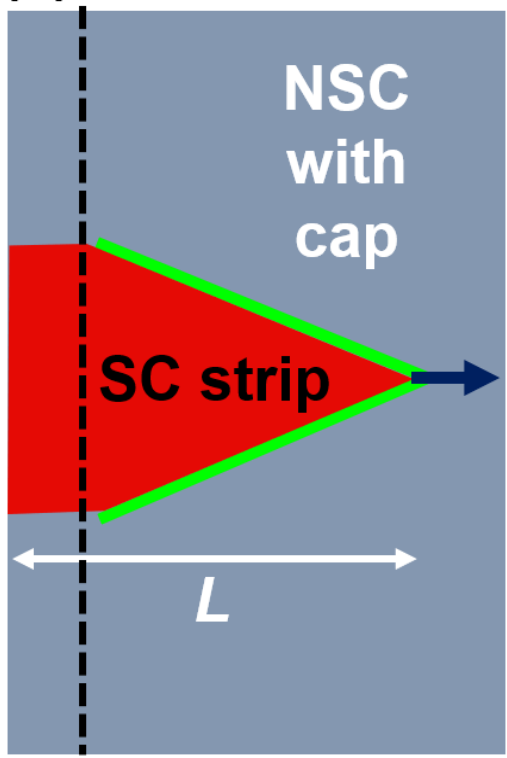

FIG2

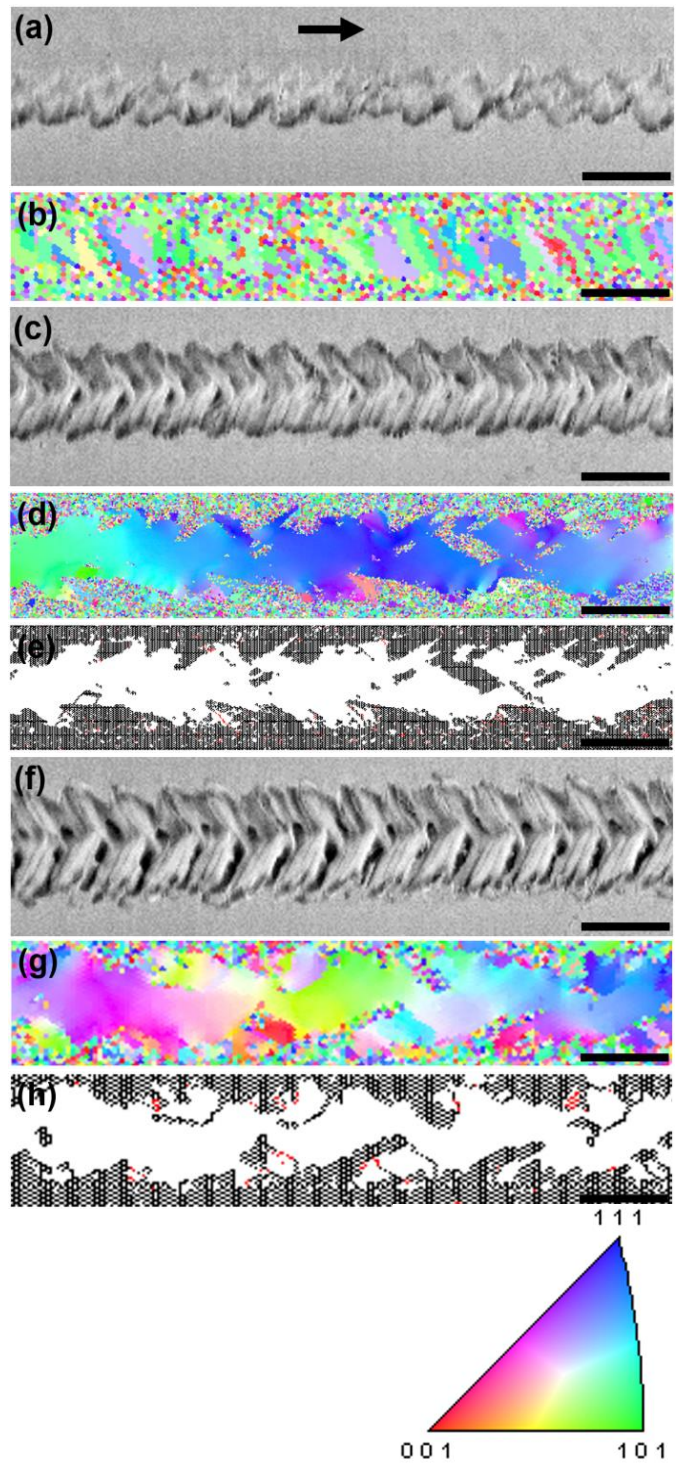

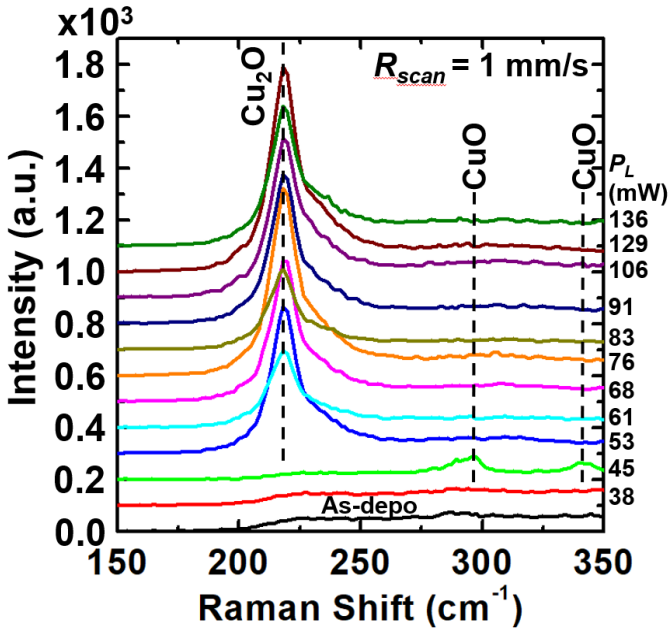

FIG4

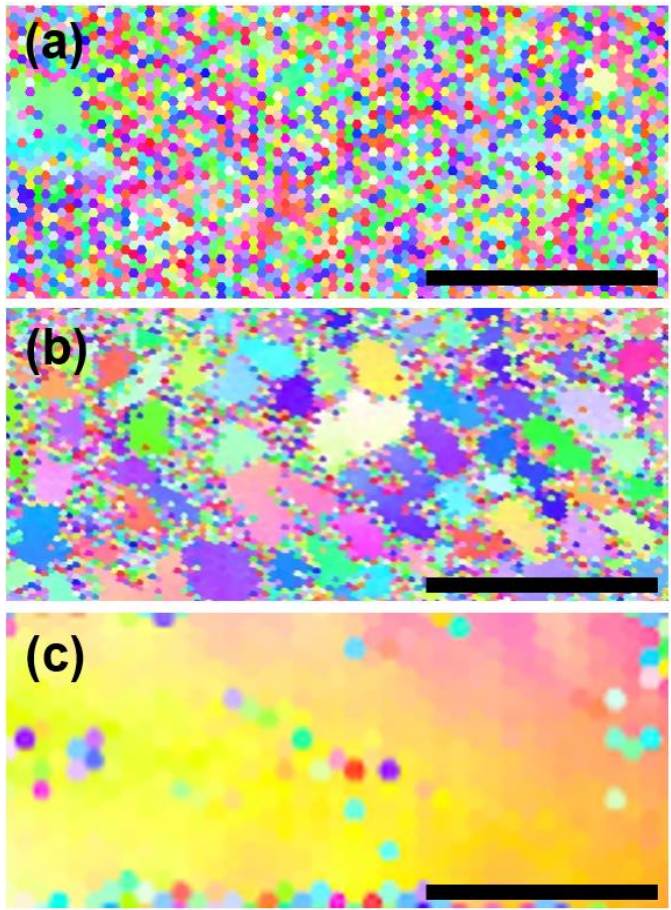

FIG5

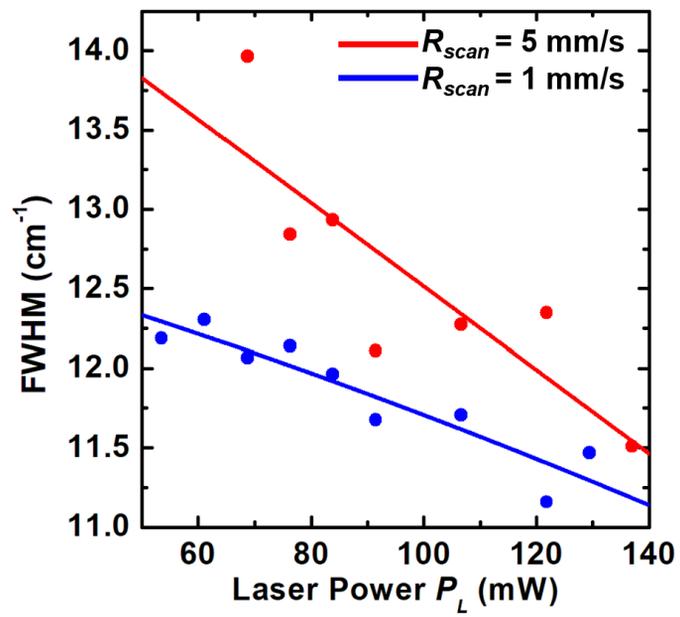

FIG6

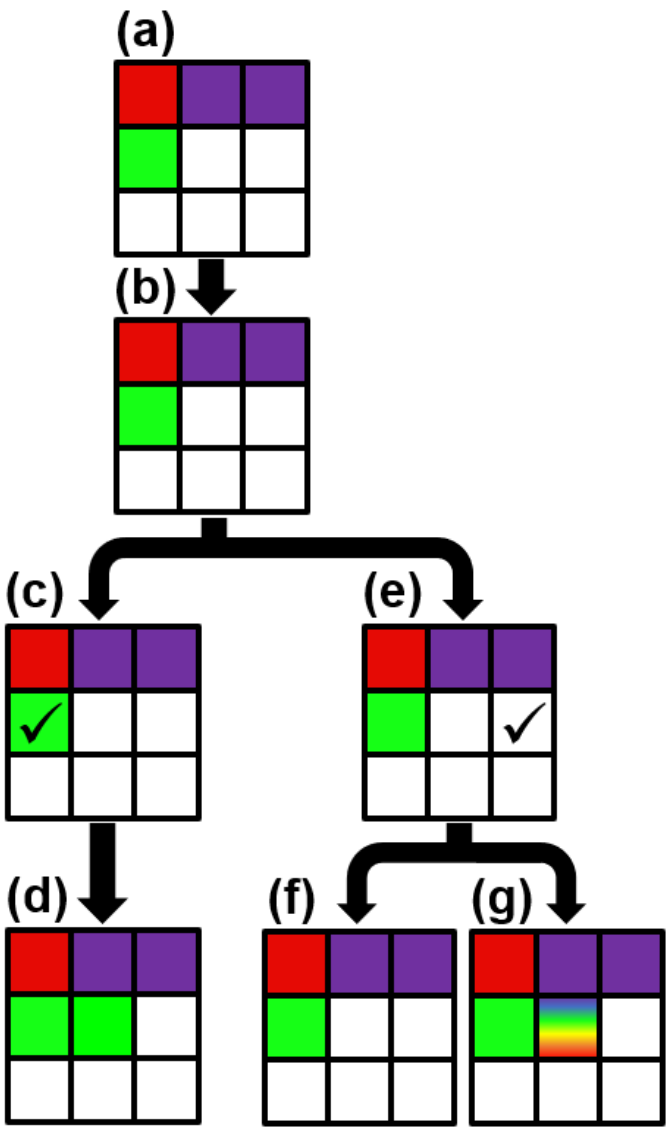

FIG7

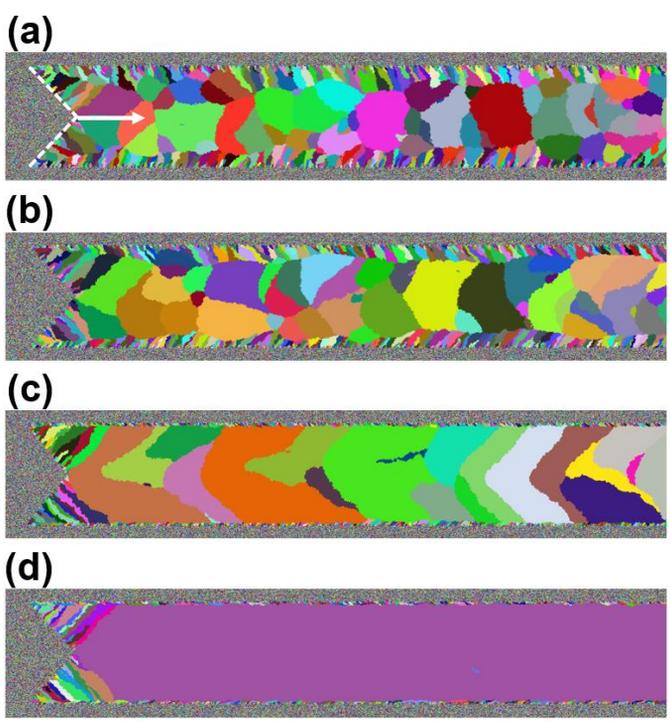

FIG8

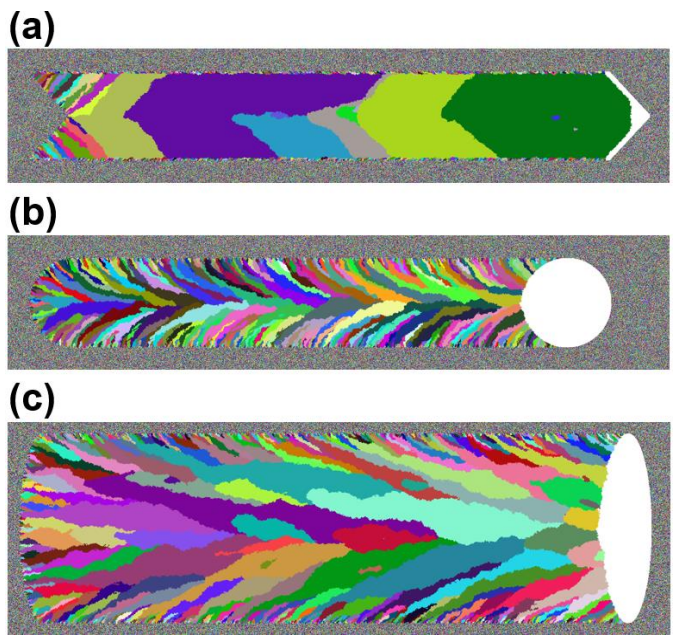

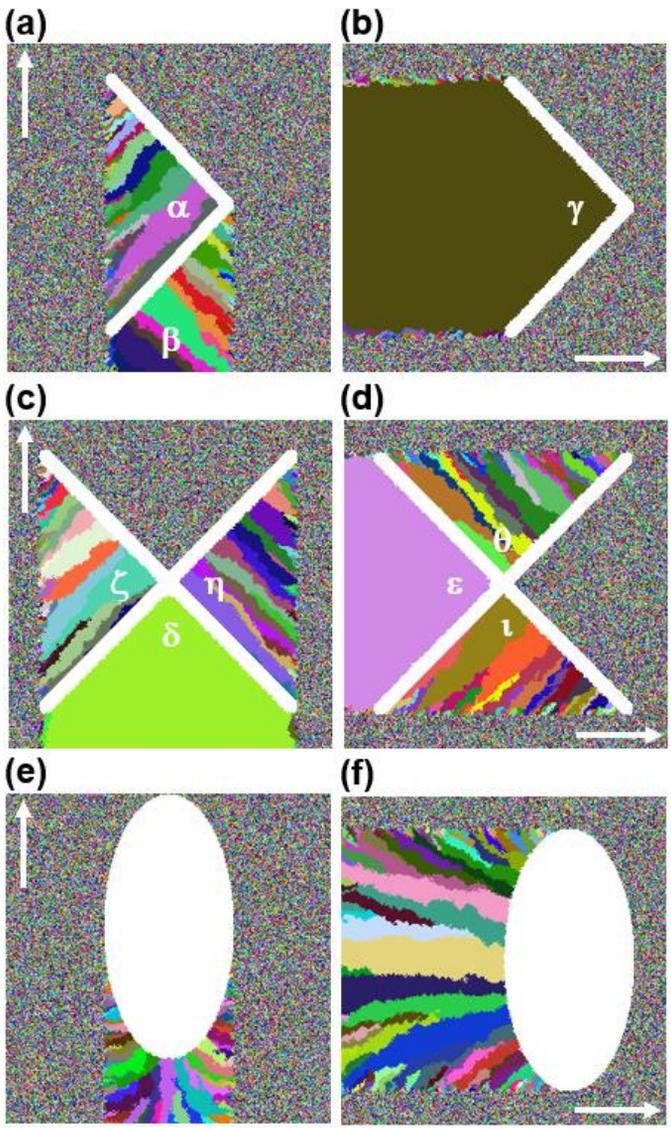
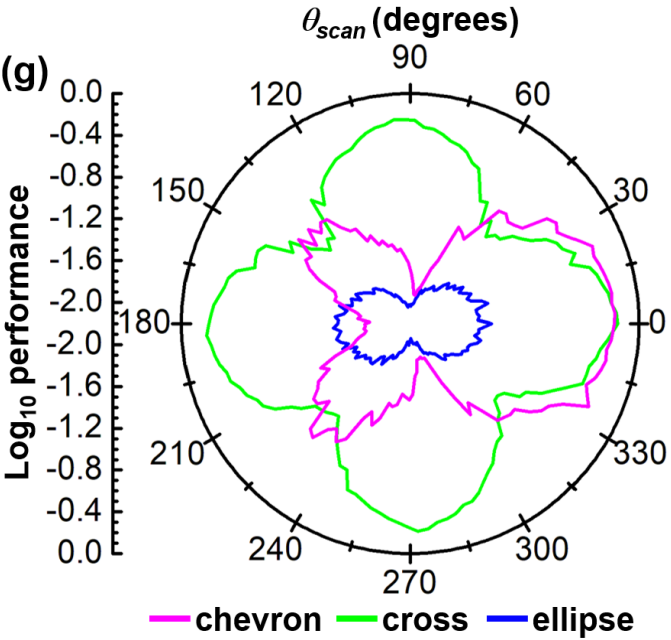

FIG10

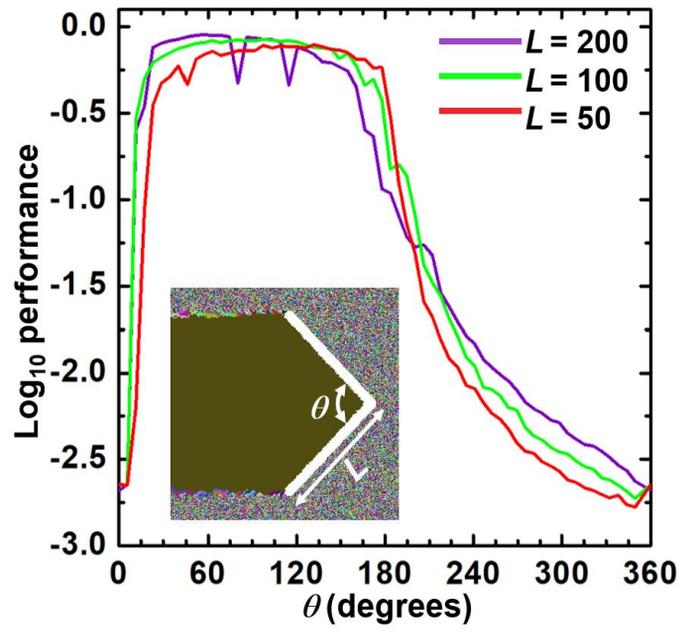